\documentstyle[prb,aps,epsf,psfig,multicol]{revtex}
\newcommand{\la}{\langle}
\newcommand{\ra}{\rangle}
\newcommand{\ua}{\uparrow}
\newcommand{\da}{\downarrow}
\newcommand{\rar}{\rightarrow}
\begin{document}
\draft

\title{Specific Heat of the Periodic Anderson Model: From Weak to Strong
Coupling}

\author{N. M. R. Peres}
\address{Departamento de F\'{\i}sica, Universidade de \'Evora,
Rua Rom\~ao Ramalho, 59, P-7000-671 \'Evora, Portugal\\
Centro de F\'{\i}sica da Universidade do Minho, Campus Gualtar, 
P-4700-320 Braga, Portugal}
\author{P. D. Sacramento}
\address{Departamento de F\'{\i}sica and  CFIF, Instituto Superior 
T\'ecnico, Av. Rovisco Pais, 1049-001 Lisboa, Portugal}
\author{M.A.N. Ara\'ujo}
\address{Departamento de F\'{\i}sica, Universidade de \'Evora,
Rua Rom\~ao Ramalho, 59, P-7000-671 \'Evora, Portugal\\
Centro de F\'{\i}sica da Universidade do Minho, Campus Gualtar, 
P-4700-320 Braga, Portugal}

\date{\today}

\maketitle

\begin{abstract}
We study the temperature dependence of the 
 specific heat in the
periodic Anderson 
model as function of the on-site Coulomb interaction,
hybridization, and position of the $f-$electrons energy level. 
At strong coupling
($U=\infty$) we use slave bosons, whereas at
weak coupling we use a simple Hartree-Fock decomposition. 
We find that both in the strong and weak coupling limits
the specific heat of the system presents a multiple peak structure
in a mean field treatment. We believe these features to be related
to band-shape effects.
The temperature evolution of the low-temperature peak position 
on the model parameters for the non-half filled non-symmetric case
is discussed. 
\end{abstract}
\vspace{0.3cm}
\pacs{PACS numbers: 75.20.Hr, 71.27.+a, 71.28.+d, 65.40.+g}
\begin{multicols}{2}

Heavy-fermion materials are compounds
containing rare earth and actinide atoms that exhibit high specific heats
and spin susceptibilities at low temperature. 
This feature is  associated with
high effective masses of quasi-particles due to strong Coulomb interactions
among the $4f-$ or $5f-$electrons.\cite{hewson}
The behavior of these materials is controlled by three parameters: 
the hybridization
$V$ between  $f-$ and $d-$ orbitals, which confers the $f-$electrons
an itinerant character, the value of the $f-$level $\epsilon_0$
energy  relatively
to the chemical potencial, and the value of the Coulomb energy $U$.
\cite{newns87} 
 
The strong  Coulomb interaction 
together with the small values of the hybridization matrix allow for formation
of local moments leading to magnetic behavior in these systems. 
The Kondo effect and the magnetic order due to the RKKY
(Ruderman-Kittel-Kasuya-Yosida) 
interaction
are consequences of the interplay between Coulomb
interaction and hybridization.\cite{millis87,houghton88}  
Together with magnetism, heavy-fermion materials may present
superconducting ground states, whose origin and nature are far
from being fully understood.
\cite{bernhoeft98,also00,estrela00} 

The periodic Anderson model (PAM) is considered 
a good candidate for 
describing heavy-fermion systems. 
It is generally accepted that a complete understanding of 
heavy-fermion system properties requires going beyond mean field theory.
Nevertheless, mean field theory still gives qualitatively 
correct information about some properties of heavy-fermion systems.
\cite{newns87} 

Recently, the temperature dependence of the 
specific heat of the PAM, at half filling, 
was studied using 
the equation of motion
method, both in the symmetric and 
non-symmetric cases,\cite{lacroix99,luo00} and 
a multiple peak structure was found. 
This multiple peak structure was first obtained by
Zieli\'nski and Matlak in the half filled symmetric ($2\epsilon_0+U=0$) 
case.\cite{zielinski82}
However, it is believed that mean field theory is not able to
capture this multiple peak structure in the specific heat.

This belief contrasts with Monte Carlo results 
showing that the magnetic Hartree-Fock approximation reproduces well
the free energy of the PAM at low
temperatures when $V\ll U$,\cite{carey99}
implying that the thermodynamics must be well described by the Hartree-Fock
approximation in this limit.

In this study we report results for the  PAM's specific heat  both in
the weak and infinite coupling limits.
Our study is done away from half filling
and in the non-symmetric case. 
The strong-coupling limit $U=\infty$ is studied using slave bosons,
whereas the weak coupling limit
is treated within Hartree-Fock approximation. 
We find that in both limits the
specific heat presents a multiple peak structure and that its dependence
on the model parameters is in qualitative agreement with what is
physically expected. 


The standard  periodic Anderson model
model for spin 1/2 electrons is written as

\begin{equation}
H=H^0_d+H_f^0+H_{df}+H_U\,,
\label{hi}	
\end{equation}
where
\begin{eqnarray}
H_d^0&=&\sum_{\vec k,\sigma}(\epsilon_{\vec k}-\mu)
d_{\vec k,\sigma}^{\dag}d_{\vec k,\sigma}\,,\\
H_f^0&=&\sum_{i,\sigma}(\epsilon_0-\mu)f_{i,\sigma}^{\dag}f_{i,\sigma}\,,\\
H_{df}&=&V\sum_{i,\sigma}\left(d_{i,\sigma}^{\dag}f_{i,\sigma}
+f_{i,\sigma}^{\dag}d_{i,\sigma}\right)
\label{hcf}\,,\\
H_U&=&U\sum_{i}n_{i,\uparrow}^f n_{i,\downarrow}^f\,,
\end{eqnarray}
and $\epsilon_{\vec k}$ is the dispersion of the $d-$electrons,
$\epsilon_0$ is the bare energy of the 
localized $f-$states, $\mu$ is the chemical potential, $V$
is the hybridization matrix element (assumed $\vec k$ independent),
$U$ is the on-site Coulomb interaction, 
and $n_{i,\sigma}^f=f_{i,\sigma}^{\dag}f_{i,\sigma}$.
This model describes two electronic subsystems, the $d$ and $f-$electron
systems, which are coupled by the hybridization matrix. Since the
$f-$states form very narrow bands, the Coulomb interaction
has a strong effect on the $f-$states. 

It is well known that, at the mean field level, the PAM presents a 
two-band structure, separated by a gap. Therefore, 
the model describes a Kondo-insulator for  $n=2$, describing a metal
for other values of $n$. The full density of states shows a sharp
peak of width $V^2\rho_0$ 
(with $\rho_0$ the $d-$electron density of states)
close to the renormalized $f-$ level energy, which is responsible
for the quasiparticle heavy-masses when the chemical potential
lies within this peak. 

In our study we consider the non-magnetic phase such that 
$\la n_{\ua}^f\ra=\la n_{\da}^f\ra$. 
In the strong coupling limit ($U=\infty$),
double occupancy of the $f$ sites is forbidden.
The simplest
implementation of  the $U=\infty$  condition  is  due 
to Coleman,\cite{coleman84} in which the  empty $f$-site is represented
by a slave boson $b_i$ and  the physical 
operator $f_i$ in equation  (\ref{hcf})
is replaced with  ${b^{\dagger}}_i f_i$. 
Condensation of the slave-bosons can be described
by the replacement  $b_i \rightarrow \la b_i\ra=\la{b^{\dagger}}_i\ra
=\sqrt{z}$.  

The  density of the  boson condensate $z$  minimizes
the free energy of the system and  the
renormalized $\epsilon_f$ energy level is obtained 
after imposing
local particle (boson+fermion) conservation at the  $f$-sites.
Using the results of Ref. 
\onlinecite{peres00} 
for the non-superconducting case,
the mean field equations can be written
in terms of the Fourier transform of the Green's functions 
\begin{eqnarray}
{\cal G}_{f,\sigma}(\vec k,\tau-\tau') & = & 
	-\la T_{\tau}f_{\vec k,\sigma}(\tau)
f^{\dag}_{\vec k,\sigma}(\tau')\ra\,,\\
{\cal G}_{d,\sigma}(\vec k,\tau-\tau') & = & 
	-\la T_{\tau}d_{\vec k,\sigma}(\tau)
d^{\dag}_{\vec k,\sigma}(\tau')\ra\,,\\
{\cal G}_{df,\sigma}(\vec k,\tau-\tau') & = & 
	-\la T_{\tau}d_{\vec k,\sigma}(\tau)
f^{\dag}_{\vec k,\sigma}(\tau')\ra\,,
\end{eqnarray}
as

\begin{eqnarray}
z  &=&  1- \frac{T}{N_s} \sum_{\vec{k},\sigma} \sum_{i\omega_n}
 {\cal G}_{f,\sigma} (\vec{k},
i\omega_n)\nonumber\\
&=& 1-2\int_{-D}^Dd\epsilon\,\rho_0(\epsilon)[u^2_+(\epsilon)f(E_+)+
 u^2_-(\epsilon)f(E_-)] 
\label{z}\,,
\end{eqnarray}
and 
\begin{eqnarray}
\epsilon_f   &=&\epsilon_0  - \frac{VT}{\sqrt{z}N_s}
 \sum_{\vec{k}, \sigma} \sum_{i
\omega_n} {\cal G}_{df,\sigma} (\vec{k}, i\omega_n)\nonumber\\
  &=&\epsilon_0-\frac{2V}{\sqrt z}
\int_{-D}^Dd\epsilon\,\rho_0(\epsilon)
\sqrt{u^2(\epsilon)_+u^2_-(\epsilon)}\times\nonumber\\
&&[f(E_+)-f(E_-)]\,,
\label{novadelta}
\end{eqnarray}
where $N_s$ denotes the number of lattice sites,  
$\epsilon_f$ is the renormalized
energy of the $f$ orbitals due to the on-site repulsion
and $f(E)$ is the Fermi function.
Equation (\ref{z}) states that the  mean number of 
electrons at an $f$-site is $1-z$.
For a given number of particles per site, $n$, these
equations  must be supplemented with the  particle conservation condition 
which yields the chemical potential $\mu$ for any temperature:
\begin{eqnarray}
n &=&  1-z+\frac{T}{N_s} \sum_{\vec{k},\sigma} \sum_{i\omega_n}
 {\cal G}_{d,\sigma} (\vec{k},
i\omega_n)\nonumber\\
&=&2\int_{-D}^Dd\epsilon\,\rho_0(\epsilon)
[f(E_+)+f(E_-)]
\,.
\label{mean}
\end{eqnarray}
In the above equations  the energies $E_{\pm}$ and 
the coherence factors $u^2_{\pm}(\epsilon)$ 
are given by
\begin{eqnarray}
E_{\pm}&=&\frac 1 2 (\epsilon_{\vec k}+\epsilon_f)\pm \frac 1 2 
E(\epsilon_{\vec k})\,,\nonumber\\ 
E(\epsilon_{\vec k})&=&\sqrt{(\epsilon_{\vec k}-\epsilon_f)^2+4zV^2}\,,\nonumber\\
u^2_{\pm}(\epsilon)&=& \frac 1 2 \left[
1 \mp \frac {\epsilon_{\vec k}-\epsilon_f}{E(\epsilon_{\vec k})}\right]\,.
\end{eqnarray}
and $\rho_0(\epsilon)$ represents the density of states of the  free 
$d-$electrons, which we choose of the form
\begin{equation}
\rho_0({\epsilon})=\frac {2}{\pi D^2}\sqrt{D^2-\epsilon^2}\,,\hspace{0.5cm}
-D\le \epsilon \le D\,,
\end{equation}
where $D$ is half the $d$-electrons bandwidth. 

At weak coupling the on-site Coulomb repulsion
term $H_U$ is treated using the usual Hartree-Fock decomposition
$
H_U\rar  U  \la n^f\ra/2\sum_{i,\sigma}n^f_{i,\sigma}\,.
$
Since the total number of electrons, $n_d$+$n_f$, is fixed, we obtain
a mean field equation for $n_f$, which is  solved
together with the equation for the chemical potential $\mu$. The Coulomb
interaction contributes to the renormalization of $\epsilon_0$, leading
to a renormalized $f-$ energy given by 
$
\epsilon_f=\epsilon_0+ U  \la n^f\ra/2\,. 
$

Once the mean field equations are solved, the specific heat is computed as
\begin{equation}
C=\frac {1}{N_s}\frac{d\, <H>}{d\,T}\,.
\end{equation}
At weak coupling, the mean field parameters are $n_f$ and $\mu$,
whereas at $U=\infty$ these are  $n_f$, $\epsilon_f$  and $\mu$.
In what follows we take $D=6$ and $U$, $\epsilon_0$, $V$ and $T$ 
are measured in units of $D$.

\begin{figure}
\begin{center}
\epsfxsize=6cm
\epsfbox{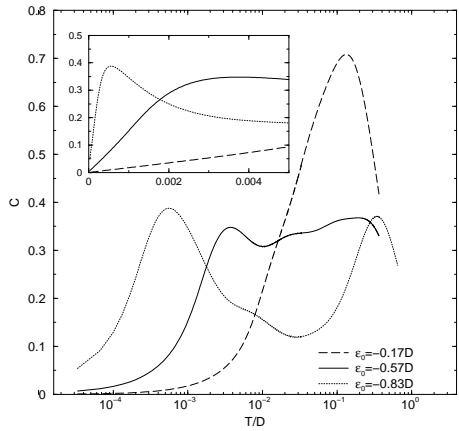}
\end{center}
\caption{Specific heat of
the PAM in the $U=\infty$ limit as function of $T/D$ for different values of 
$\epsilon_0$. The inset shows a linear low-temperature 
behavior of $C(T)$, as it should be for a metal.
The parameters are  $V=0.17D$, $n=1.1$.
}
\label{sbshe0}
\end{figure}

\begin{figure}
\begin{center}
\epsfxsize=6cm
\epsfbox{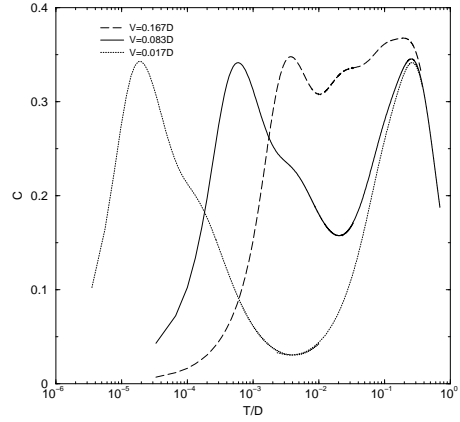}
\end{center}
\caption{
Specific heat  as function of $T/D$ for different values of 
$V$, in the $U=\infty$ regime. 
The parameters are $\epsilon_0=-0.57D$ and  $n=1.1$.}
\label{sbshV}
\end{figure}

\begin{figure}
\begin{center}
\epsfxsize=7cm
\epsfbox{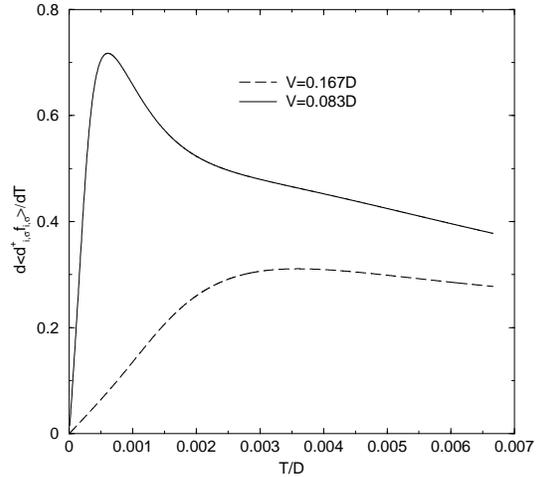}
\end{center}
\caption{Temperature dependence of
the temperature derivative of matrix element
$\la d^\dag_{i,\sigma}f_{i,\sigma}$ 
for two different values of $V$,
in the $U=\infty$ regime, for $\epsilon_0=-0.57D$ and $n=1.1$.}
\label{sbdtV}
\end{figure}

\begin{figure}
\begin{center}
\epsfxsize=7cm
\epsfbox{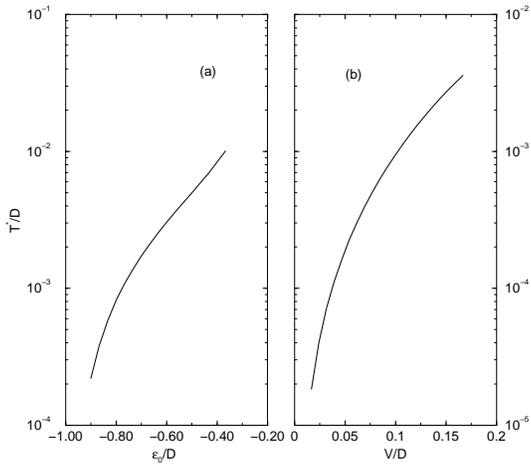}
\end{center}
\caption{Dependence of the correlation temperature
$T^{\ast}$ as a function of: (a) $\epsilon_0$ and (b) $V$
in the $U=\infty$ regime and  for $n=1.1$.}
\label{sbtkV}
\end{figure}

In Figures \ref{sbshe0} and 
\ref{sbshV} we show the specific heat of the PAM in the
$U=\infty$ limit for several values of 
 $\epsilon_0$ and $V$, respectively. Since we are in the 
metallic regime the specific heat is linear at low temperatures
(check inset in Fig. \ref{sbshe0}). Both the $\gamma$ coefficient 
and the zero-temperature spin susceptibility are much
higher than  the corrresponding values for non-hybridized $d$-electrons,
signaling the presence of heavy quasi-particles. 
We see that  as $V$  
or  $\epsilon_0$ decrease (with $ \epsilon_0<0$) 
the low temperature peak is shifted to lower
and lower temperatures, whereas the temperature position 
of the high temperature peak remains essentially constant. 
Let us define a characteristic temperature $T^{\ast}$
as the temperature value
where $d\la d^{\dag}_{i,\sigma}f_{i,\sigma}\ra /d\,T$ has its first 
local  maximum, as can be seen in Fig. \ref{sbdtV}
for two values of $V$. 
A similar definition has been adopted for the Kondo temperature $T_K$ 
(on the lattice) in
Ref. \onlinecite{lacroix99}. Comparing the results for $T^\ast$  
in  Fig. \ref{sbtkV} with those in Figs. \ref{sbshe0} and 
\ref{sbshV} for $C(T)$,  
we see that the temperature position of the low-temperature 
peak coincides with $T^{\ast}$. 
The dependence of $T^\ast$
on $V$ and $\epsilon_0$ 
follows the same trends as those found  for $T_K$ in
Ref. \onlinecite{lacroix99},
for the half-filled symmetric case,
where a Kondo intra-site interaction of the
form $J_K\sim V^2U/(\epsilon_0(\epsilon_0+U))$,
was assumed to be present.
Since we have not included the fluctuations of the boson
fields,
it is remarkable that 
mean field theory (in the boson fields) can capture
this effect
since the Kondo interactions are not  generated
at mean field level.  We interpret the appearence of
a specific heat peak at temperature $T^*$ as being 
associated with the existence of a  peak  in the density of states.

\begin{figure}[f]
\begin{center}
\epsfxsize=6cm
\epsfbox{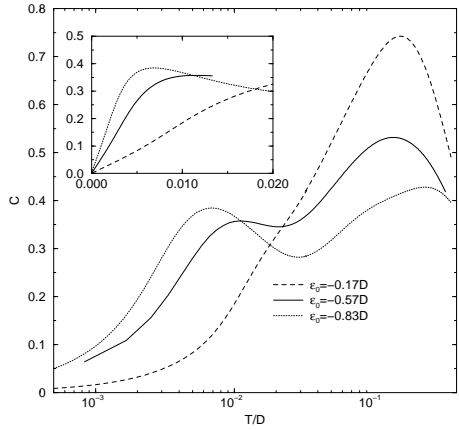}
\end{center}
\caption{Specific heat of the
PAM in the Hartree Fock approximation 
 as function of $T/D$ for different values of 
$\epsilon_0$. The inset shows a linear low-temperature 
behavior of $C(T)$, as it should be for a metal.
The parameters are $U=0.47D$, $V=0.17D$, $n=1.1$.}
\label{hfshe0}
\end{figure}

\begin{figure}[f]
\begin{center}
\epsfxsize=6cm
\epsfbox{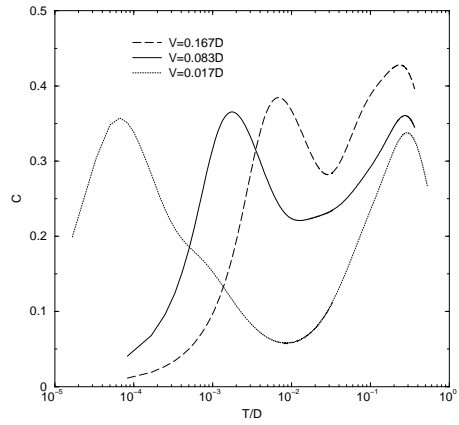}
\end{center}
\caption{Specific heat 
of the
PAM in the Hartree Fock approximation
as function of $T/D$ for different values of 
$V$.
The parameters are $U=0.47D$, $\epsilon_0=-0.83D$, $n=1.1$.}
\label{hfshV}
\end{figure}

As we have seen, in the strong coupling limit $U=\infty$ the constraint
$n_f\le 1$ introduces boson fields. At the mean field level, the effect
of these fields is to renormalize the hybridization matrix $V$
to $V_z=\sqrt z V$. This
implies that for the same parameters as in the weak coupling
(Hartree-Fock) regime
the low temperature peak in $C(T)$ shifts to lower temperatures.
This can be clearly seen comparing Fig. \ref{sbshe0} with
Fig.  \ref{hfshe0},
and Fig. \ref{sbshV} with Fig. \ref{hfshV}. 
We have
also checked that the low-temperature peaks for 
different $V$ or $\epsilon_0$ tend to collapse into
a single peak if $T$ is scaled by $T^{\ast}$, as observed for the
symmetric case in Ref. \onlinecite{lacroix99}. It is also clear
from Figs. \ref{sbshe0} and \ref{sbshV} that as we move the system
towards $n_f\rar 1$, which in a more elaborated description
would correspond to the Kondo limit,  
the single peak structure separates into 
two peaks. It is also clear that the low-temperature peak
is in fact a superposition of two peaks not completely separated. 
In Ref. \onlinecite{lacroix99} the complete separation of these
two peaks took place only at very negative values of $\epsilon_0$.

In the single impurity case the model has been solved by Bethe Ansatz.
\cite{Okiji} In that case, $C(T)$ also presents a double peak structure.
In that solution the charge and spin excitations are described by
different rapidities becoming clear
that the low-temperature peak in $C(T)$
is associated with the spin excitations
whereas the high-temperature peak is due to the charge excitations. 
 An exact solution is not available for the lattice problem, however.

In more sophisticated  treatments  of the PAM
\cite{lacroix99,luo00} a lattice Kondo temperature
$T_K$ has been identified with the first maximum in the
specific heat  leading the authors to  conclude that the first peak is
associated with the Kondo shielding (spin excitations) whereas
the second peak is associated with the charge transport (at an energy
of the order of the hopping taken as unity in this paper).
But the calculation we just presented shows
 that both the slave-boson mean field
theory and the most simple Hartree Fock approximation
lead to a  multiple peak structure in the specific heat.
These approximations do not describe the dynamics or screening
 of local moments.
 We  therefore conclude that the double peak structure in the 
specific heat
is simply a band-shape effect, namely, the appearence of
a peak in the density of states. Indeed, such a feature is also
present  in more sophisticated treatments of the PAM.\cite{lacroix99,luo00}

If the system undergoes
a magnetic or superconducting transition,
the peak structure is replaced, in general, 
 by a lambda-shaped peak signalling the transition to the 
ordered phase or phases if they coexist. 
In the particular case of a superconducting ground state \cite{peres00}
and zero magnetization we have observed that in some cases the
low-temperature peak is  still present. A study of the ordered phases
will be presented elsewhere.\cite{npm}

In many heavy-fermion materials the $f-$electrons have large total angular
momentum. 
At mean field level and for $N>2$ 
the $z$ parameter has a critical temperature given by 
$T_z=(\epsilon_f-\mu)/\ln(N-1)$. In the ''Kondo regime'' 
$\epsilon_f-\mu$ is
a very small quantity.\cite{millis87,coleman84} 
As the temperature rises $V_z$  vanishes.
Therefore, we expect for $C(T)$, and within mean field theory,
a different temperature behavior
from that reported here.

In summary, we have found that $C(T)$ for the PAM  presents a double
peak structure both in the strong coupling limit and in the
Hartree-Fock approach. This multiple peak structure in $C(T)$
is associated with a peak in the density of states of the system,
due to hibridization between delocalized $d-$ and localized $f-$
states.
The temperature  $T^{\ast}$ 
of the first peak corresponds to a maximum
in the  derivative of $\la d^{\dag}_{i,\sigma}f_{i,\sigma}\ra$ with
respect to temperature.
$T^\ast$ decreases as both $V$ and $\epsilon_0$
decrease. This effect is much more pronounced in the strong coupling
regime, since $V$ is renormalized to smaller values due to the factor
 $\sqrt{z}$. 


This research was supported by the Portuguese Program
PRAXIS XXI under grant number 2/2.1/FIS/302/94.


\end{multicols}

\end{document}